\documentclass[preprint]{aastex62}

\usepackage{mathtools}

\submitjournal{ApJL}

\shorttitle{Red Spirals}
\shortauthors{Hao et al.}

\begin{document}

\title{Spatially Resolved Studies of Local Massive Red Spiral Galaxies}

\correspondingauthor{Cai-Na Hao}
\email{hcn@bao.ac.cn}

\author[0000-0002-0901-9328]{Cai-Na Hao}
\affiliation{Tianjin Astrophysics Center, Tianjin Normal University, Tianjin 300387, China}

\author[0000-0002-8614-6275]{Yong Shi}
\affiliation{School of Astronomy and Space Science, Nanjing University, Nanjing 210093, China}
\affiliation{Key Laboratory of Modern Astronomy and Astrophysics (Nanjing University), Ministry of Education, Nanjing 210093, China}

\author{Yanmei Chen}
\affiliation{School of Astronomy and Space Science, Nanjing University, Nanjing 210093, China}
\affil{Key Laboratory of Modern Astronomy and Astrophysics (Nanjing University), Ministry of Education, Nanjing 210093, China}

\author{Xiaoyang Xia}
\affiliation{Tianjin Astrophysics Center, Tianjin Normal University, Tianjin 300387, China}

\author{Qiusheng Gu}
\affiliation{School of Astronomy and Space Science, Nanjing University, Nanjing 210093, China}
\affiliation{Key Laboratory of Modern Astronomy and Astrophysics (Nanjing University), Ministry of Education, Nanjing 210093, China}

\author{Rui Guo}
\affiliation{Tianjin Astrophysics Center, Tianjin Normal University, Tianjin 300387, China}

\author{Xiaoling Yu}
\affiliation{School of Astronomy and Space Science, Nanjing University, Nanjing 210093, China}
\affiliation{Key Laboratory of Modern Astronomy and Astrophysics (Nanjing University), Ministry of Education, Nanjing 210093, China}

\author{Songlin Li}
\affiliation{School of Astronomy and Space Science, Nanjing University, Nanjing 210093, China}
\affiliation{Key Laboratory of Modern Astronomy and Astrophysics (Nanjing University), Ministry of Education, Nanjing 210093, China}

\begin{abstract}

We report two-dimensional spectroscopic analysis of massive red spiral
galaxies  ($M_{*}$ $>$  10$^{10.5}$ $M_{\odot}$)  and compare  them to
blue  spiral and  red elliptical  galaxies above  the same  mass limit
based on  the public SDSS  DR15 MaNGA  observations. We find  that the
stellar population properties of red  spiral galaxies are more similar
to those  of elliptical  galaxies than to  blue spiral  galaxies.  Red
spiral galaxies  show a  shallow mass-weighted  age profile,  and they
have  higher stellar  metallicity and  Mgb/${\rm \langle  Fe \rangle}$
across the whole  1.5$R_{\rm e}$ as compared to blue  spirals, but all
these  properties  are close  to  those  of elliptical  galaxies.  One
scenario to explain this is that red spirals form as remnants of very
gas-rich major mergers that happened above $z$$\sim$1.

\end{abstract}




\section{Introduction} \label{sec:intro}

The global properties of galaxies show bimodality in both optical colors and
morphologies, and these two properties are closely linked to each other: spiral
galaxies are mostly blue and elliptical galaxies appear red \citep{Baldry04,
Bell04, Schawinski2014}. It is generally believed that a spiral galaxy grows
its disk inside-out through a gradual mode by long-standing gas accretion and
moderate star formation rates \citep{Mo98, Munoz-Mateos07}, and a massive
elliptical galaxy is produced through an intense mode in which merging of two spiral
galaxies triggers strong starbursts and transforms morphologies violently
\citep{Hopkins06}.  However, the existence of red passive spiral galaxies, those
with spiral morphologies but red colors (little star formation), poses a
challenge to the above popular scenarios \citep{Bundy10, Masters10}.  They
could be a special phase during secular evolution in which star formation has
already ceased but morphologies stay intact due to the lack of violent events,
such as major mergers \citep{Masters10}. They could also be products of some
other physical processes, e.g., galaxy merging with high gas fractions that
instead produces spiral galaxies \citep{Springel05, Robertson06, Robertson08,
Hopkins09, Athanassoula16, Sparre17}. 

Many efforts have been devoted to understanding the origin of red
spiral galaxies \citep[e.g.,][]{Skibba09,Robaina12}.  Existing investigations
of red spirals revealed some distinct properties as compared to normal blue
spirals \citep{Bundy10, Masters10,Tojeiro13}: their star formation rates and
dust contents are low with old central stellar populations as expected from
their red colors; they reside in all kinds of environments with preference in
intermediate local densities; they have higher fractions of Seyferts and LINERs
(Low Ionization Nuclear Emission Regions) and significantly higher bar
fractions; their stellar light distributions are more concentrated. However,
investigations of spectroscopic properties of their stellar populations are
limited.

In Guo et al. (2019, submitted, hereafter Paper I), we selected a red massive
spiral galaxy sample from the Sloan Digital Sky Survey Data Release 7 (SDSS
DR7) and compared their central and global properties with those of blue spiral
and red elliptical galaxies based on the SDSS fiber spectra and photometric
data.  We found that red spirals are more similar to red ellipticals than to
blue spirals of comparable masses in their central stellar populations, stellar
mass surface densities and host dark matter halo masses.  These results suggest
that the central parts of red spiral galaxies formed at a similar epoch and
timescale to those of red ellipticals. Our Paper I and previous studies from
other groups \citep{Masters10,Robaina12,Tojeiro13} are mainly based on the
fiber spectra from SDSS, which can only sample the bulge component of local red
spiral galaxies.  It is essential to extend similar studies to disk components
to have a more complete census of their stellar populations.

In this paper, we explore the two-dimensional spectroscopic data as observed by
the survey of the Mapping Nearby Galaxies at Apache Point Observatory (MaNGA)
\citep{Bundy15, Drory2015, Law15, Law2016, Yan2016a, Yan2016b, Blanton2017,
Wake2017}. Such data enable constraints on the spatial distributions of
kinematics and stellar population properties such as age, metallicity and
$\alpha$-element enhancement, which will provide important clues to the origin
of red spiral galaxies. We adopt a flat $\Lambda$CDM cosmology with
$\Omega_{\rm m}=0.3$, $\Omega_{\rm \Lambda}=0.7$, $H_{\rm 0}=70\,{\rm km \,
s^{-1} Mpc^{-1}}$.

\section{Sample Selection And MaNGA Data} \label{sec:data}

\begin{figure*}[tbh]
  \begin{center}
    \includegraphics[scale=0.5]{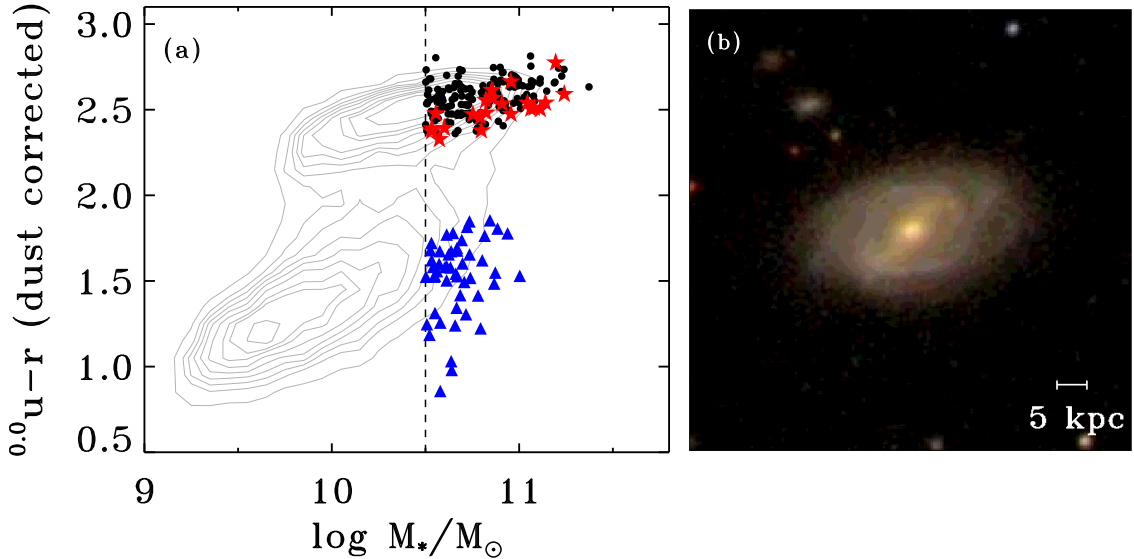}
    \caption{\label{sample} {\bf Selections of samples}. {\bf (a)}, massive red spiral galaxies (red symbols) are selected
      using the $u-r$ color vs. stellar mass diagram from the SDSS, along with blue spirals (blue symbols)
      and red ellipticals (black symbols). The contours show the number density distribution of galaxies
in the redshift range of $0.02 < z < 0.05$ and absolute $z-$band magnitude
of $M_{z,\rm Petro} < -19.5$ in the catalog of \citet{Mendel2014}. The dashed line marked the stellar mass of 10$^{10.5}$
      M$_{\odot}$. {\bf (b)}, an SDSS false-color image of a red spiral galaxy.}
\end{center}
\end{figure*}

Our sample of massive red spiral galaxies with stellar masses above 10$^{10.5}$
$M_{\odot}$ is selected from the SDSS DR7 \citep{Abazajian09}, with their
dust-corrected $u-r$ color in the red sequence (see Figure~\ref{sample}  (a))
and spiral features in their morphologies (see Figure~\ref{sample} (b)).  To
better understand their origin, we compare red spirals to blue spirals and red
ellipticals above the same mass limit, as shown in Figure~\ref{sample} (a).
The sample selection is detailed in Paper I. Briefly, galaxies with redshift of
$0.02 <  z  < 0.05$ and stellar masses above 10$^{10.5}M_\odot$ were selected
from the catalog of \citet{Mendel2014}. The morphological information from the
Galaxy Zoo 1 project \citep{Lintott11} was then adopted to divide the sample
into spirals and ellipticals. To  minimize the impact of the dust attenuation,
spirals with disc inclination angle  $i > 60\deg$ were rejected.  Then spirals
and ellipticals were separated into blue and red according to their locations
in the dust-corrected $u-r$ color versus stellar mass diagram following similar
criteria in \citet{Guo2016} (see Figure~\ref{sample} (a)).  After
cross-matching these galaxies with the data release 15 of the SDSS for the
MaNGA \citep{Bundy15, Law15, Aguado2019},  we were left with 22 red spirals,
along with 50 blue spirals and 164 red ellipticals, plotted in
Figure~\ref{sample} (a).  As can be seen from this figure, red spirals and
ellipticals span a wider range in mass than blue spirals towards the high mass
end. Considering possible effects of the mass dependence of the stellar
population and kinematic properties, we extracted subsamples by restricting the
stellar masses to the range of $10^{10.5} - 10^{11}M_\odot$, which yields 15
red spirals, 49 blue spirals and 139 ellipticals. We will study both the
samples with $M_* > 10^{10.5}M_\odot$ (referred to ``the main samples'',
hereafter) to keep the sample sizes as large as possible and the subsamples
with $10^{10.5} < M_* < 10^{11}M_\odot$ to minimize the mass dependence effect.
Since only one blue spiral galaxy was removed from the main sample to form the
subsample, we do not expect any changes in their properties. We note that our
samples of galaxies are not complete samples but they should be representative
of the respective types of galaxies. 

The MaNGA data analysis pipeline \citep{Westfall2019}, which uses pPXF
\citep{Cappellari04, Cappellari2017} and the MIUSCAT \citep{Vazdekis12} stellar
library, fits the stellar continuum in each spaxel and produces estimates of
the stellar kinematics and lick indices, including $V_{\rm star}$, $\sigma_{\rm
star}$ and  Mgb/${\rm \langle Fe \rangle}$ (=Mgb/(0.5*Fe5270+0.5*Fe5335)) used
in this study.  The stellar metallicity and stellar age are taken from the
MaNGA Pipe3D value added catalog \citep{Sanchez16, Sanchez2018} included in the
SDSS DR 15. To investigate the formation of the bulk of the stars in galaxies,
we use the stellar mass-weighted ages and metallicities in this work.  Median
radial profiles of the above quantities will be used to characterize the
properties of each type of galaxies. They are the azimuthally averaged radial
profiles of galaxies within each category. To better understand the properties
of the bulge and disk components for spiral galaxies, we adopt the effective
radius of the bulge component as a measure of the bulge size, which was derived
by Sersic+Exponential fitting to the 2D surface brightness profile of the SDSS
$r$-band image \citep{Fischer2019}. The median bulge sizes for red and blue
spirals are around\footnote{The main samples and the subsamples show slightly
different median bulge sizes.} 0.15\,$R_{\rm e}$ and 0.3\,$R_{\rm e}$,
respectively, where $R_{\rm e}$ is the effective radius of the total light
distribution.

\section{RESULTS} \label{sec:results}

\begin{figure*}[tbh]
  \begin{center}
    \includegraphics[scale=0.6]{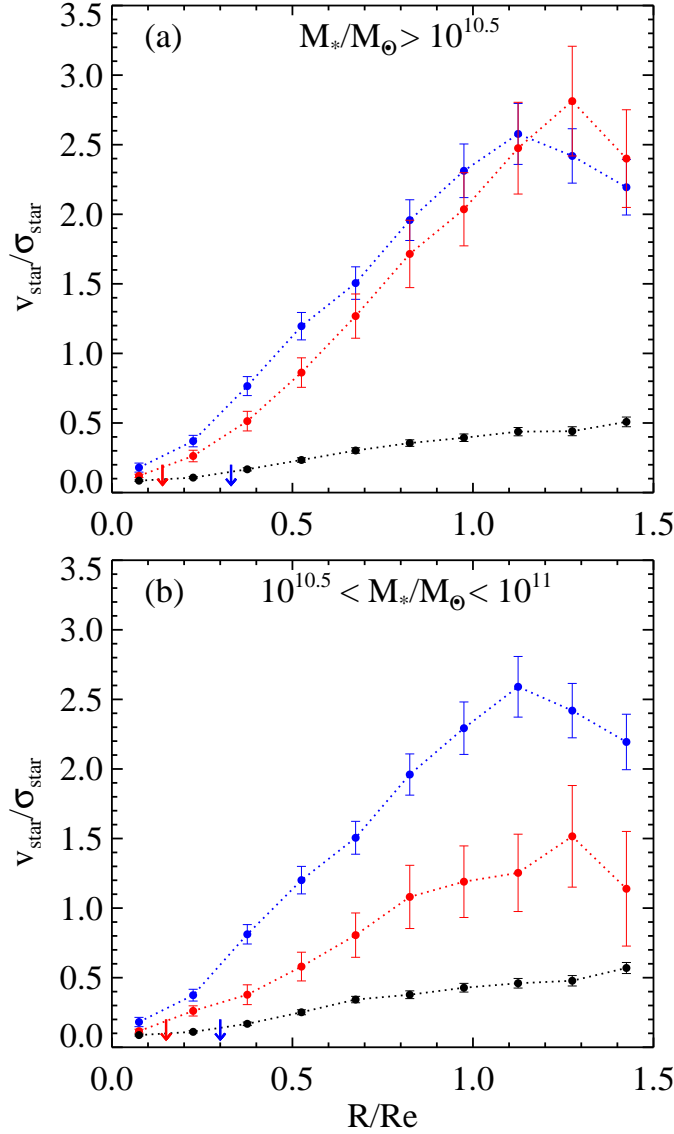}
\caption{\label{vel_vdisprat}{\bf The median radial profiles of $V_{\rm star}/\sigma_{\rm star}$} for red spiral galaxies (red),
      blue spiral galaxies (blue) and red elliptical galaxies (black) in the main samples {\bf (a)} and in the subsamples {\bf (b)}, with a radial bin of
      0.15 $R_{\rm e}$. The error bar represents the error of the mean. The downward arrows indicate the median effective radius of the bulge component
for red spirals (red) and blue spirals (blue).}
\end{center}
\end{figure*}

We first examine the kinematics of our three types of galaxies to verify
whether the kinematic estimator of morphologies are consistent with the optical
one. Figure~\ref{vel_vdisprat} (a) shows the median radial profile of the ratio
of rotation velocity to velocity dispersion for blue spiral, red spiral and
elliptical galaxies in the main samples. The red and blue arrows indicate the
median bulge size of red spiral and blue spiral galaxies, respectively. As
shown in the figure, red spiral galaxies have similar kinematics to blue
spirals, with dispersion-dominated bulges in the inner regions plus the
rotation-dominated disks in the outer regions. Elliptical galaxies show
dispersion-dominated kinematics from the central part to the outer part.
Figure~\ref{vel_vdisprat} (a) confirms the optical morphological classification
that red spirals do harbor bulges and rotating disks. The results for the
subsamples are shown in Figure~\ref{vel_vdisprat} (b). For both ellipticals and
blue spirals, the $V_{\rm star}/\sigma_{\rm star}$ profiles stay the same as
those for the main samples. By contrast, red spirals show remarkably lower
ratios of velocity to velocity dispersion in their disks than the main sample,
which is mainly caused by the decrease in the rotational velocity as we
examined, a result of the Tully-Fisher relation \citep{Tully1977}. But they
are still consistent with being systems harboring bulges and rotating disks.

\begin{figure*}[tbh]
  \begin{center}
    \includegraphics[scale=0.7]{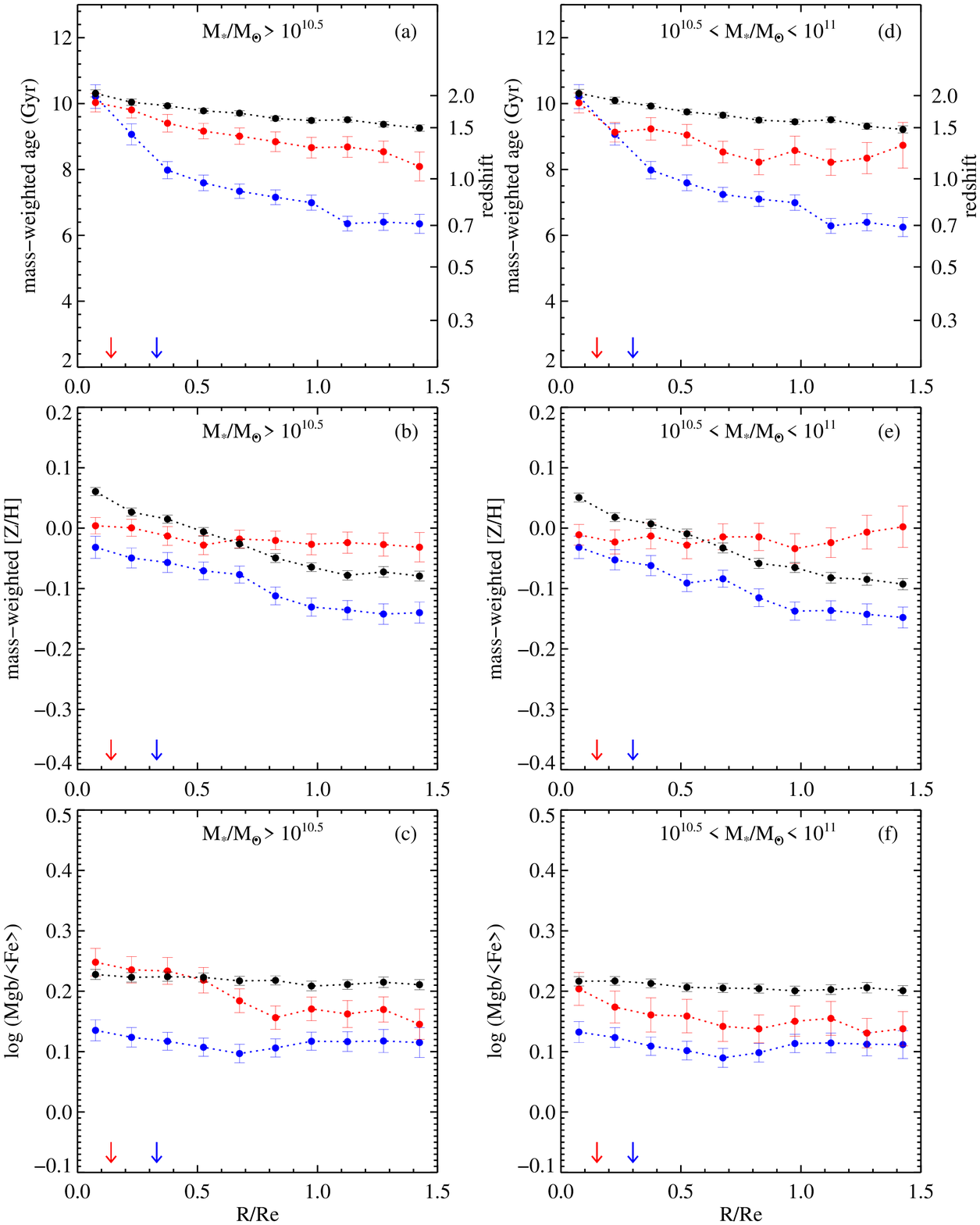}
    \caption{\label{radial_profile} {\bf  The median radial profiles of
      stellar mass-weighted age (a), stellar mass-weighted metallicity  (b),
      and  log(Mgb/${\rm \langle Fe \rangle}$) (c)} for red spiral galaxies (red),
      blue spiral galaxies (blue) and red elliptical galaxies (black) in the main samples.
      The same properties for the subsamples are shown in {\bf (d)-(f)}. 
      The data points connected by dotted lines show the profiles along the ellipse
      with   position   angle   and   ellipticity   from   NASA-Sloan
      Atlas \citep{Abazajian09,  Blanton11}.
      The bin sizes are 0.15 $R_{\rm e}$. The error bar represents the error of the mean.
      The downward arrows indicate the median effective radius of the bulge component
for red spirals (red) and blue spirals (blue).}
\end{center}
\end{figure*}

In Figure~\ref{radial_profile}, we present the median radial profiles of the
stellar population properties  of the three types of galaxies including stellar
ages, metallicities and Mgb/${\rm   \langle  Fe \rangle}$.
Figure~\ref{radial_profile} (a)-(c) are for the main samples of galaxies. As
shown in Figure~\ref{radial_profile} (a), red spirals and elliptical  galaxies
show similar shallow age profiles out to 1.5  $R_{\rm e}$, although red spirals
are slightly younger than ellipticals. As labeled in the y-axis of the figure,
red spirals formed their bulge and disk components within $\sim1.5$\,Gyr before
redshift $\sim1.2-1.3$, and ellipticals formed within $\sim1$\,Gyr  before
redshift $\sim1.5$.   In contrast, the central region ($<$ 0.3  $R_{\rm e}$) of
blue  spirals shows a steep mass-weighted  age gradient  with  similar ages to
red  spirals  and ellipticals at the very center,  while their disk components
show flat age gradients but are much younger than red spirals and ellipticals.
The above results suggest that red spirals most likely harbor star formation
histories that resemble ellipticals instead of blue spirals.  Blue spirals
started to form their bulges at a similar time to red spirals but have an
extended star formation history in the outer parts, which is consistent with a
prolonged inside-out formation scenario.

Figure~\ref{radial_profile} (b) presents the median radial profiles of the
stellar metallicities. Red spirals show a rather flat metallicity profile from
the center out to 1.5 $R_{\rm e}$. Blue spiral galaxies and ellipticals show
similarly steeper metallicity gradients than red spirals, with blue spirals
being more metal-poor than ellipticals. Compared to red spirals, blue spirals
are more metal-poor over the entire probed radius, and the metallicity
difference varies from $\sim0.03$ dex to $\sim0.1$ dex. On the other hand,
elliptical galaxies are more metal-rich in the inner region ($\lesssim 0.5-0.6
R_{\rm e}$) but more metal-poor in the outer region ($\gtrsim 0.5-0.6 R_{\rm
e}$) than red spirals.

Figure~\ref{radial_profile} (c)  shows the  median radial  profiles of
Mgb/${\rm  \langle Fe  \rangle}$. Mgb  is used  to probe  the $\alpha$ elements
\citep[e.g.,][]{Worthey92,Zheng2019}  that are  produced  by type II supernova,
while Fe is synthesized in type Ia supernova. The $\alpha$/Fe ratio  traces the
relative importance  of  the  intense starburst  and  the subsequent long-term
quiescent  star  formation \citep{Matteucci86, Worthey92, Thomas05}.  As  shown
in the figure, the Mgb/${\rm  \langle Fe \rangle}$ profile  of red spirals
breaks  into three parts: flat profiles in the inner ($\lesssim 0.5 R_{\rm e}$)
and outer regions ($\gtrsim 0.8 R_{\rm e}$), and an intermediate declining
profile  connecting the  higher log(Mgb/${\rm  \langle Fe  \rangle}$) values
($\sim 0.23$)  in the inner region and  the lower log(Mgb/${\rm \langle Fe
\rangle}$) values ($\sim  0.16$) in the outer region.  In comparison,
elliptical galaxies show a  very flat Mgb/${\rm \langle Fe \rangle}$  profile
and  high log(Mgb/${\rm  \langle  Fe  \rangle}$) values about $0.22$.  Blue
spirals also have a  flat Mgb/${\rm \langle Fe \rangle}$  profile but their
log(Mgb/${\rm  \langle Fe \rangle}$) values  are relatively  low  ($\sim
0.12$).   The  higher  values  of Mgb/${\rm \langle Fe \rangle}$ for elliptical
galaxies suggest a short star formation timescale while the lower values  of
Mgb/${\rm \langle Fe \rangle}$  for blue spiral galaxies reflect the
contribution of a more  prolonged  star formation  history.  The  Mgb/${\rm
\langle  Fe \rangle}$ values in the bulge and the inner region of the disk of
red spiral galaxies are comparable  to those of elliptical galaxies, implying
similarly short star formation timescales in the inner region of red  spiral
galaxies and elliptical galaxies.  On the  other hand, the intermediate
Mgb/${\rm \langle Fe \rangle}$  values in  the disk components of red  spiral
galaxies suggest a star formation timescale between those of ellipticals and
blue spirals.

Similar to Figure~\ref{radial_profile} (a)-(c), Figure~\ref{radial_profile}
(d)-(f) show the median radial profiles of stellar ages, metallicities and
Mgb/${\rm   \langle  Fe \rangle}$ but for the subsamples of galaxies with
stellar masses $10^{10.5} < M_* < 10^{11}M_\odot$. It is obvious that this
further stellar mass cut at the high mass end does not change the results for
ellipticals and blue spirals. Even for red spirals, both the profiles of the
stellar ages and metallicities are similar to the main samples. The only
notable change is in the Mgb/${\rm \langle  Fe \rangle}$ profile of red spirals
at $R \lesssim 0.7 R_{\rm e}$.  Red spirals in the subsample show lower
Mgb/${\rm   \langle  Fe \rangle}$ values than those in the main sample, which
suggests that more massive red spirals have higher Mgb/${\rm   \langle  Fe
\rangle}$ and hence shorter star formation timescales in their inner regions.
In spite of this decrease in the Mgb/${\rm   \langle  Fe \rangle}$ values, they
are still larger than blue spirals. This suggests that the star formation
timescales of red spirals are in between the ellipticals and blue spirals
except the most central (bulge) region that shows an identical Mgb/${\rm
\langle  Fe \rangle}$ value with the ellipticals.  In summary, the subsample of
red spirals shows some different properties from the main sample in their disk
components, but this does not affect the relative difference between these
three types of galaxies.

\section{Discussion}\label{sec:discussion}

The stellar population properties presented in Figure~\ref{radial_profile} for
blue spiral galaxies and elliptical galaxies are in good agreement with those
shown in the literature \citep[e.g.,][]{Gonzalez2015,Zheng2017}. Blue spiral
galaxies have an extended star formation history with their bulge components
formed earlier than their disk components, consistent with the inside-out
formation scenario as induced by long-standing gas accretion and associated
star formation. Elliptical galaxies show old stellar ages and enhanced
Mgb/${\rm \langle Fe \rangle}$ values, in agreement with the early formation
epoch and short star formation timescales. 

While studies of morphologies and kinematics of red spirals indicate that they
harbor typical spiral disks in addition to the  central bulges, their stellar
populations are instead found to be distinctly different from blue spirals: red
spirals have an older and shallower mass-weighted age profile than blue
spirals, with their disks younger than their bulges by $\sim 1-2$\,Gyr; they
show a more metal-rich and flatter mass-weighted metallicity profile relative
to blue spirals; their Mgb/${\rm \langle Fe  \rangle}$ values are in between
ellipticals and blue spirals, being enhanced compared to blue spirals.

The above results suggest that red spiral galaxies are not evolutionary
remnants of blue spiral galaxies. Instead, they are consistent with the remnant
of galaxy merging with high gas fractions. While in general galaxy merging
produces elliptical galaxies \citep{Hopkins06}, theoretical simulations
predicted that merging with high gas fractions \citep[f$_{\rm gas} > 0.5$ at
the time of merger; e.g.,][]{Robertson06} instead produces spiral galaxies
\citep{Springel05, Robertson06, Hopkins09, Athanassoula16, Sparre17}.  During
such a very gas-rich major merger, gravitational torques make gas within some
characteristic radius lose angular momentum and fall into the center rapidly,
and then form stars in a starburst mode \citep{Hopkins09}. This process leads
to the formation of a bulge component with high metallicity and Mgb/${\rm
\langle Fe \rangle}$.  On the other hand, gas at sufficiently large radii that
cannot fall in efficiently, or gas that cannot efficiently dissipate or lose
the angular momentum will cool quickly and re-form a rotating disk.
Subsequently, the gas in the disk form stars but with much reduced star
formation rates \citep{Springel05}, which will result in a new stellar disk
with younger age and lower Mgb/${\rm \langle Fe \rangle}$ than the bulge formed
earlier in the starburst. The disk can still have similar metallicity to the
bulge assuming the galaxy is quenched soon after formation.  All these
theoretical expectations are in good agreement with our observational results
shown in Figure~\ref{radial_profile}. 

Furthermore, we note that very gas-rich major merging events may only happen at
relatively high redshifts (above $z\sim$1) when the gas content was rich. This
condition could be satisfied, as shown by the old stellar populations of red
spirals in Figure~\ref{radial_profile}(a).  If the proposed scenario is
plausible, red spirals thus offer a window to understanding the formation of
massive spiral galaxies through gas-rich major mergers instead of the gradual
inside-out growth mode.

The most difficult part of the merging scenario is to explain the shutdown of
the star formation in red spirals. The IFU data shown here could not put any
constraints on the star formation quenching mechanisms. But the investigations
on the bulge to total stellar mass ratios and the halo masses in our paper I
provide some hints on possible quenching mechanisms. Red spiral galaxies show
higher bulge to total stellar mass ratios than blue spiral galaxies and more
than 80\% of red spiral galaxies are hosted by dark matter halos heavier than
$10^{12}M_\odot$, above which halo quenching takes effect. So morphological
quenching \citep{Martig2009} may be responsible for the cease of the star
formation and the massive dark matter halos may have been preventing gas from
cooling and falling onto the galaxies to form new stars.

\section{Conclusion} 

We carried out two-dimensional spectroscopic analysis of massive red spiral
galaxies with $M_{*}$ $>$ 10$^{10.5}$ $M_{\odot}$ out to 1.5$R_{\rm e}$,  and
compared them to blue spiral and red elliptical galaxies above the same mass
limit based on the SDSS DR 15 for the MaNGA survey. We found that the stellar
population properties of red spiral galaxies are more similar to those of red
elliptical galaxies than to blue spiral galaxies. They have higher stellar
metallicity and Mgb/${\rm \langle Fe \rangle}$ across the whole 1.5$R_{\rm e}$
as compared to blue spirals, and their stars are old with relatively flat
mass-weighted age gradients in contrast to those of blue spiral galaxies.
These results prove that red spiral galaxies are not evolutionary remnants of
blue spiral galaxies.  They are possible remnants of gas-rich major mergers
above $z\sim1$ instead.  Major mergers of disk galaxies with sufficiently high
gas content can form a bulge rapidly in the first place and then form a gas
disk. The gas in the disk will form stars subsequently.

\acknowledgments

We thank the anonymous referee for constructive comments that improved the
paper.  We would like to thank Drs. Cheng Li, Shude Mao, Dandan Xu, Zheng
Zheng, Fangzhou Jiang, and Jian Fu for helpful discussions. This work is
supported by the National Key Research and Development Program of China (No.
2017YFA0402703) and the National Natural Science Foundation of China (NSFC, No.
11733002 and 11373027). Y.S. also acknowledges the support from the National
Key R\&D Program  of  China  (No.  2018YFA0404502, No. 2017YFA0402704), the
NSFC (No.  11825302, 11773013)  and the Excellent Youth Foundation of the
Jiangsu Scientific  Committee (BK20150014).  Y.C. also acknowledges the support
from the National Key R\&D Program of China (No. 2017YFA0402704) and the NSFC
(No.  11573013).  Funding for the Sloan Digital Sky Survey IV has been provided
by the Alfred P. Sloan Foundation, the U.S. Department of Energy Office of
Science, and the Participating Institutions. SDSS-IV acknowledges support and
resources from the Center for High-Performance Computing at the University of
Utah. The SDSS web site is www.sdss.org.  SDSS-IV is managed by the
Astrophysical Research Consortium for the Participating Institutions of the
SDSS Collaboration including the Brazilian Participation Group, the Carnegie
Institution for Science, Carnegie Mellon University, the Chilean Participation
Group, the French Participation Group, Harvard-Smithsonian Center for
Astrophysics, Instituto de Astrof\'isica de Canarias, The Johns Hopkins
University, Kavli Institute for the Physics and Mathematics of the Universe
(IPMU) / University of Tokyo, the Korean Participation Group, Lawrence Berkeley
National Laboratory, Leibniz Institut f\"ur Astrophysik Potsdam (AIP),
Max-Planck-Institut f\"ur Astronomie (MPIA Heidelberg), Max-Planck-Institut
f\"ur Astrophysik (MPA Garching), Max-Planck-Institut f\"ur Extraterrestrische
Physik (MPE), National Astronomical Observatories of China, New Mexico State
University, New York University, University of Notre Dame, Observat\'ario
Nacional / MCTI, The Ohio State University, Pennsylvania State University,
Shanghai Astronomical Observatory, United Kingdom Participation Group,
Universidad Nacional Aut\'onoma de M\'exico, University of Arizona, University
of Colorado Boulder, University of Oxford, University of Portsmouth, University
of Utah, University of Virginia, University of Washington, University of
Wisconsin, Vanderbilt University, and Yale University.  This project makes use
of the MaNGA-Pipe3D data products. We thank the IA-UNAM MaNGA team for creating
this catalogue, and the ConaCyt-180125 project for supporting them.

\end{document}